\documentclass[journal]{IEEEtran}
\usepackage[T1]{fontenc} 
\usepackage{cite}
\usepackage{subfigure}
\usepackage{graphicx}
\usepackage{amsmath}
\usepackage{mathtools}
\usepackage{comment}
\usepackage{enumitem}
\usepackage{booktabs}
\usepackage{multirow} 
\usepackage{epstopdf}
\usepackage{array}
\usepackage{wrapfig}
\usepackage[flushleft]{threeparttable}

\newcommand{\innermid}{\nonscript\;\delimsize\vert\nonscript\;}
\newcommand{\activatebar}{
  \begingroup\lccode`\~=`\|
  \lowercase{\endgroup\let~}\innermid 
  \mathcode`|=\string"8000
}

\usepackage{amsthm,amssymb}
\usepackage{setspace}
\newcommand{\subparagraph}{}
\usepackage[compact]{titlesec}
\usepackage{hyperref}
\hypersetup{
    colorlinks=true,
    linkcolor=blue,
    filecolor=magenta,      
    urlcolor=cyan,
}

\usepackage{bm}
\usepackage{tikz}
\usetikzlibrary{shapes,arrows}

\begin{document}

\tikzstyle{decision} = [diamond, draw, fill=blue!20, 
    text width=4.5em, text badly centered, node distance=3cm, inner sep=0pt]
\tikzstyle{block} = [rectangle, draw, fill=blue!20, 
    text width=5em, text centered, rounded corners, minimum height=4em]
\tikzstyle{line} = [draw, -latex']
\tikzstyle{cloud} = [draw, ellipse,fill=red!20, node distance=3cm,
    minimum height=2em]

\title{Predicting Encoded Picture Quality in Two Steps is a Better Way}
\author{Xiangxu Yu, Christos G. Bampis, Praful Gupta and Alan C. Bovik
\thanks{X. Yu, C. G. Bampis, P. Gupta and A. C. Bovik are with the Department of Electrical and Computer Engineering, University of Texas at Austin, Austin, USA (e-mail: yuxiangxu@utexas.edu; bampis@utexas.edu; praful\textunderscore gupta@utexas.edu; bovik@ece.utexas.edu).}
}
\maketitle

\begin{abstract}
Full-reference (FR) image quality assessment (IQA) models assume a high quality "pristine" image as a reference against which to measure perceptual image quality. In many applications, however, the assumption that the reference image is of high quality may be untrue, leading to incorrect perceptual quality predictions. To address this, we propose a new two-step image quality prediction approach which integrates both no-reference (NR) and full-reference perceptual quality measurements into the quality prediction process. The no-reference module accounts for the possibly imperfect quality of the source (reference) image, while the full-reference component measures the quality differences between the source image and its possibly further distorted version. A simple, yet very efficient, multiplication step fuses the two sources of information into a reliable objective prediction score. We evaluated our two-step approach on a recently designed subjective image database and achieved standout performance compared to full-reference approaches, especially when the reference images were of low quality. The proposed approach is made publicly available at https://github.com/xiangxuyu/2stepQA.

\end{abstract}


\IEEEpeerreviewmaketitle

\section{Introduction}
\IEEEPARstart{T}\ \hspace{-0.89mm}\MakeLowercase{h}e past few years have witnessed tremendous growth in the viewing and sharing of digital images \cite{wang2006modern}. Numerous consumer-driven and social media applications, such as Snapchat, Facebook, and Twitter allow users to rapidly post and share images that they capture with their handheld devices. These images are prone to a number of very common in-capture distortions, such as camera noise, poor exposure, motion blur, and compression artifacts. In this letter, we only consider still pictures that undergo two stages of distortions (source impairments followed by compression), but the principle is extensible to videos. Streaming companies such as Netflix and Youtube offer a large amount of legacy content including old movies or television programs \cite{bovik201775}. These video streams, which are obtained from diverse sources, are often subjected to source-related artifacts such as interlacing, film grain or upscaling. These source videos of lower quality are then encoded for transmission. Being able to evaluate the quality of these twice-distorted images and videos on the receiver side is of great interest.

In the majority of cases, human viewers are the end users, and picture quality evaluation is crucial when storing, processing and transmitting visual data. Since subjective quality rating is impractical at scale, numerous perceptually-designed objective image quality assessment (IQA) algorithms have been developed for image quality prediction. These are broadly classified either as (Full or Reduced) Reference models, if both the distorted image and ostensibly pristine version of it are both available to be compared. Full reference models such as SSIM \cite{wang2004image}, MS-SSIM \cite{wang2003multiscale}, FSIM \cite{zhang2011fsim} and VSI \cite{zhang2014vsi} deliver excellent quality prediction performance on subjective quality databases such as LIVE IQA \cite{sheikh2006statistical}, TID2008 \cite{ponomarenko2009tid2008} and CSIQ \cite{larson2010most}. Reduced-reference (RR) models such as RRED \cite{soundararajan2012rred} and SpEED-QA \cite{bampis2017speed} extract perceptually relevant features representing only a subset of the avaliable image information when estimating image quality. No-reference (NR) IQA algorithms are applied when only a distorted image is available. These models can be further divided into opinion- (or distortion-) aware (OA) models such as BRISQUE \cite{mittal2012no}, DIIVINE \cite{moorthy2011blind} and BLIINDS \cite{saad2012blind} which are trained on human opinion scores and/or on specific distortions, and opinion-unaware (OU) models like NIQE \cite{mittal2013making} and IL-NIQE \cite{zhang2015feature}. These OU models are "completely blind" approaches in that they are only driven by natural scene statistics (NSS) models of good quality images.

A common assumption of Reference IQA models that are used to quality-assess compressed images is that the source image to be coded is of high quality, hence can be regarded as a "pristine" reference. However, when source artifacts are present, this assumption is violated. Therefore, a Reference computation, if applied on a low-quality source image and a compressed version of it, will likely produce an incorrect quality score. Here we shown that Reference quality prediction of imperfect images that are subsequently compressed can be greatly improved by accounting for the source image quality. Towards this end we have devised a simple but effective two-step NR-R IQA concept which is applied on a pair of images: a possibly distorted reference image and a compressed (hence further distorted) version of it. An initial NR step captures the prior quality of the reference image before compression,  while the Reference step determines the further quality degradation between the reference and the distorted images. Here we focus on processing a possibly distorted reference image to predict the quality of a JPEG compressed version of it. This development is organized as follows. Section \ref{2} details our two-step IQA approach, while Section \ref{3} describes a recently developed subjective image database that we created to develop and evaluate the two-step method. Section \ref{4} discusses the experimental results and Section \ref{5} concludes with avenues for future work.

\section{Two-Step NR-FR IQA Model}
\label{2}

\begin{figure}[htp]
 \centering
 \subfigure[MOS: 79]{
   \label{fig:subfig:a}
   \includegraphics[width=0.2\textwidth]{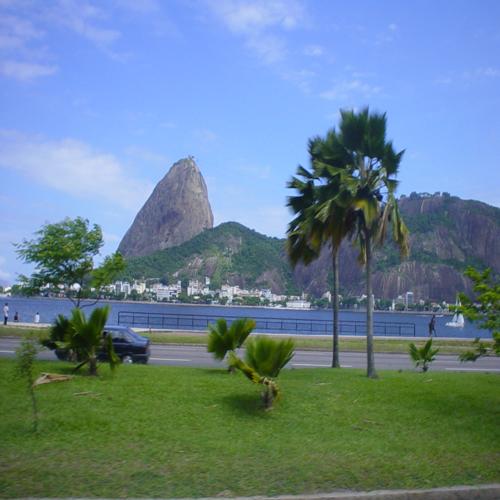}}
 \subfigure[MOS: 51\newline DMOS: 52\newline MS-SSIM: 0.9346]{
   \label{fig:subfig:b} 
   \includegraphics[width=0.2\textwidth]{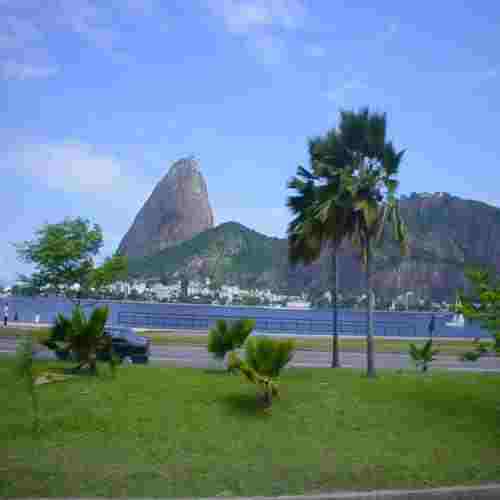}}
 \subfigure[MOS: 45]{
   \label{fig:subfig:c} 
   \includegraphics[width=0.2\textwidth]{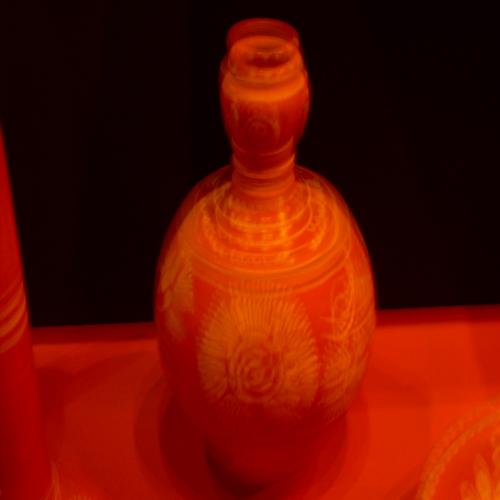}}
 \subfigure[MOS: 36\newline DMOS: 32\newline MS-SSIM: 0.9674]{
   \label{fig:subfig:d}
   \includegraphics[width=0.2\textwidth]{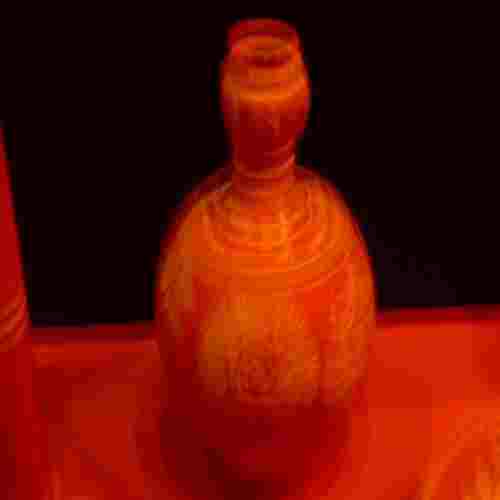}}
 \caption{(a) A high quality reference image. (b) A JPEG compressed version of (a). (c) A low quality reference image. (d) A JPEG compressed version of (c).}
 \label{fig:subfig} 
\end{figure}

Reference image quality models have generally assumed that a source reference image is of high quality, e.g., like the image shown in Fig. \ref{fig:subfig:a}. When this is true, it is  straightforward to apply a Reference model (e.g. MS-SSIM) to compare the quality of the original source image against that of its JPEG compressed version (e.g., the image in Fig. \ref{fig:subfig:b}). However, if the source is of reduced quality (Fig. \ref{fig:subfig:c}), then the Reference quality prediction may be unreliable. To illustrate this, consider the following. The source images in Fig. \ref{fig:subfig:a} and Fig. \ref{fig:subfig:c} were drawn from the LIVE In the Wild Challenge IQA database. Each has an associated mean opinion scores (MOS). In a separate study, the compressed versions in Fig. \ref{fig:subfig:b} and \ref{fig:subfig:d} were assigned both MOS and differential mean opinion scores (DMOS). MOS may be regarded as the absolute subjective quality of a distorted image, while DMOS reduces the effect of content by differencing the MOS of the source and distorted versions of each image. In practice, DMOS calculations are intended to remove content biases (e.g. due to image aesthetics), but in our case, this only reflects quality degradations arising from compression. In Fig. \ref{fig:subfig}, the MS-SSIM values are in monotonic agreement with DMOS ( increasing MS-SSIM corresponds to decreasing DMOS), while this is not the case for MOS (increases in MS-SSIM led to a decrease in MOS). This example shows that given a low quality reference image, a Reference quality prediction may fail to correctly predict the quality of a compressed version of it.

\begin{figure}
\centering
\includegraphics[width = 0.87\columnwidth]{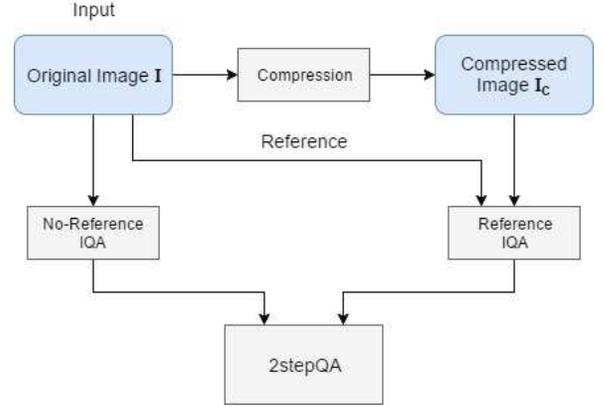}
\caption{Overview of 2stepQA.}
\label{diagram}
\end{figure}

To deal with this highly plausible situation, we introduce what we call the 2stepQA, which is a model and algorithm that combines Reference and No-Reference quality measurements in a unified manner. The concept of 2stepQA is illustrated in Fig. \ref{diagram}. Given a reference source image $\mathbf{I}$ and a compressed version $\mathbf{I_{c}}$, the Reference module compares $\mathbf{I_{c}}$ with $\mathbf{I}$, while the No-Reference module determines the quality of the source image $\mathbf{I}$. The NR prediction may be thought of as "prior" knowledge, that is used when measuring the Reference perceptual quality. The scores produced by the No-Reference and Reference modules are then combined, e.g. by computing a suitable product of the NR and FR prediction scores, yielding a final 2stepQA score.

We now visually illustrate the potential advantages of 2stepQA. Consider a hypothetical image quality axis that spans the entire quality range, e.g. from "Bad" to "Excellent". The NR block of 2stepQA evaluates the quality of $\mathbf{I}$ by, for example, computing its perceptual distance from the space of high-quality natural images. By contrast, the FR model measures the distance between $\mathbf{I}$ and $\mathbf{I_{c}}$: When $\mathbf{I}$ is of high quality, it will lie closely to the natural image space, and hence the quality of a compressed version of it may be accurately predicted as a Reference algorithm score. However, if the quality of the reference image is poor, then the result of the NR will serve to correct the Reference score. While there are many ways to combine the No-Reference and Reference scores into a single prediction, we have found that defining 2stepQA as a simple product of (functions of) the two scores to be quite effective.

\subsection{Reference IQA Module}
\label{2:FR}

A high-performance Reference IQA module is an essential ingredient of 2stepQA. An excellent choice is the multi-scale structural similarity (MS-SSIM) index \cite{wang2003multiscale}, which has found considerable commercial success. MS-SSIM extracts luminance, contrast and structural information in a multi-scale fashion, delivering quality scores ranging from 0 to 1, where larger values correspond to better quality. In our framework, MS-SSIM is applied to compare $\mathbf{I}$ with $\mathbf{I_{c}}$, the latter being degraded compression.

\subsection{No-Reference IQA Module}
\label{2:NR}

The end goal of 2stepQA is to accurately predict the perceptual quality of a compressed source image. Towards this aim, an efficient and reliable NR IQA model is also required. In our prototype 2stepQA model, we use the NIQE index \cite{mittal2013making}, which is a blind NSS-based \cite{ruderman1994statistics} OU IQA model that requires no training on distorted image or human opinions of them. The main idea behind NIQE is that the empirical distributions of mean-subtracted and divisively normalized luminance coefficients of high-quality images follow a gaussian distribution. However, in the presence of distortions, the empirical distributions tend to stray from gaussianity. By measuring these statistical deviations, a robust, blind image quality assessment engine is arrived at. Unlike data-driven approaches such as BRISQUE \cite{mittal2012no}, NIQE is very general, and serves to simplify 2stepQA while delivering good prediction performance.

\subsection{2stepQA Model}
\label{2:SCQI}

Combining Reference and No-Reference models is the main idea behind 2stepQA. As such, 2stepQA should satisfy two properties: when the reference image is of low quality, the overall 2stepQA score should reflect this, rather than relying on the Reference comparison alone. Conversely, when the reference image is of high quality, the 2stepQA score should align closely with the Reference IQA score. A direct and simple way to satisfy these properties is to define 2stepQA as a product:
\begin{equation}
Q_\mathsf{2step} = Q_\mathsf{R} \cdot Q_\mathsf{NR}
\label{2step}
\end{equation}
where $Q_\mathsf{R}$ corresponds to a suitably mapped score (here we use MS-SSIM) computed on a reference image and a distorted version of it, while $Q_\mathsf{NR}$ is a suitably mapped NR score (here, NIQE) of the reference image. A key aspect of 2stepQA is that other combinations of Reference and No-Reference models may be used. This lends a high degree of versatility to 2stepQA as provides a general framework for addressing this problem. We define the basic 2stepQA model to be
\begin{equation}
Q_\mathsf{2step} = \text{MS-SSIM} \cdot (1-\frac{\text{NIQE}}{\alpha})
\label{MN}
\end{equation}
i.e. $Q_\mathsf{R} = \text{MS-SSIM}$, $Q_\mathsf{NR} = 1-\frac{\text{NIQE}}{\alpha}$ and $\alpha = 100$. The form of $Q_\mathsf{NR}$ is arrived at since NIQE increases with worsening picture quality on a scale of about [0,100] on known databases. MS-SSIM rises with better picture quality over the range [0,1]. While 2stepQA is very simple, we will show that it is a very effective formulation.

\subsection{Data-Dependent Models}
\label{2:DD}

Other Reference and No-Reference models can be adapted in like manner. Since the many possible models deliver widely diverse ranges of quality scores, which are often data-dependent \cite{moorthy2011visual} or are trained on specific databases. In such situations, a simple linear rescaling process to match the ranges and trends of the Reference and No-Reference scores may be deployed. Specifically, rescale $Q_\mathsf{R}$ to fall in the range [0,1], where 1 represents the best possible quality. $Q_\mathsf{NR}$ is rescaled to the range [$\beta$,1], where $\beta$ $\in$ [0,1). While taking $\beta = 0$ is a possibility, we have observed that most picture quality predictors, when tested on existing databases, generate scores that cluster towards higher qualities. Thus, assume that $Q_\mathsf{R}$ is monotonic (increasing or decreasing), and that the best and worst possible quality scores delivered by the Reference engine are $R_\mathsf{hi}$ and $R_\mathsf{low}$ (either $R_\mathsf{hi}$ > $R_\mathsf{low}$ or $R_\mathsf{hi}$ < $R_\mathsf{low}$ is possible). Likewise assume that $Q_\mathsf{NR}$ is monotonic with best/worst possible quality scores $NR_\mathsf{hi}$ and $NR_\mathsf{low}$. The values of these extremal scores might be predetermined by the mathematical form of the R/NR quality models, or by measured ranges of the R/NR quality models on a database of distorted pictures. They are then rescaled as follows:
\begin{equation}
\begin{split}
&{Q}'_\mathsf{R} = a_{1}*Q_\mathsf{R}+b_{1}  \\
&{Q}'_\mathsf{NR} = a_{2}*Q_\mathsf{NR}+b_{2}
\end{split}
\label{3ab}
\end{equation}
Where $a_{1}$, $b_{1}$, $a_{2}$ and $b_{2}$ are determined using following equations:
\begin{equation}
\begin{split}
&a_{1}*R_{hi}+b_{1}=1  \\
&a_{1}*R_{low}+b_{1}=0 
\end{split}
\label{4R}
\end{equation}
\begin{equation}
\begin{split}
&a_{2}*NR_{hi}+b_{2}=1  \\
&a_{2}*NR_{low}+b_{2}=\beta 
\end{split}
\label{5NR}
\end{equation}
The constants $R_\mathsf{hi}$, $R_\mathsf{low}$, $NR_\mathsf{hi}$ and $NR_\mathsf{low}$ are specific to a pair of Reference and No-Reference models, but can be calculated offline. In our simulations, $\beta$ was determined using grid-search to find the value yielding the best performance on each database. Once computed ${Q}'_\mathsf{R}$ and ${Q}'_\mathsf{NR}$ can be substituted for $Q_\mathsf{R}$ and $Q_\mathsf{NR}$ in (\ref{2step}).

\section{A New Distorted-then-Compressed Image Database}
\label{3}
The majority of publicly available large-scale benchmark IQA databases, such as LIVE IQA \cite{sheikh2006statistical}, TID2008 \cite{ponomarenko2009tid2008} and CSIQ \cite{larson2010most}, use high quality images as reference that correspond to distorted images that were created by introducing synthetic distortions on the high quality source images. While these popular IQA databases have been quite valuable in the development of classical IQA algorithms, they do not address the degraded source quality in the context of subsequent compression distortion. We have created such a resource, which we call the LIVE Wild Compressed Picture Database. We started with a set of 80 distorted images randomly drawn from the existing LIVE In the Wild Challenge Image Quality Database \cite{ghadiyaram2016massive}, which contains a large number of images with widely diverse authentic image distortions, that were captured using a representative variety of mobile devices. 

Since our goal is to predict visual quality when images of varying quality are subjected to compression, these 80 source images were then JPEG compressed into four different and perceptually distinguishable levels, yielding four compressed images per distorted content, and 320 distorted-then-compressed images in total. We then conducted a subjective study to collect human opinion scores on the 400 images (including the 80 distorted images). The raw data was processed based on \cite{seshadrinathan2010study} and we followed the ITU-R BT 500.13 recommendation \cite{ITU} for subject rejection.

\section{Performance Evaluation}
\label{4}

\begin{table}[]
\centering
\begin{threeparttable}
\centering
\caption{Performance of Different Reference and No-Reference IQA Models on the LIVE Wild Compressed Picture Database.}
\label{alone}
\begin{tabular}{|c|c|c|}
\hline
                    & SROCC              & PCC               \\ \hline
PSNR                & 0.4227             & 0.4299            \\ \hline
MS-SSIM             & 0.8930             & 0.8923            \\ \hline
FSIM                & 0.9101             & 0.9134            \\ \hline
VSI                 & 0.7953             & 0.8153            \\ \hline
NIQE                & 0.8457             & 0.8407            \\ \hline
BRISQUE             & 0.9091             & 0.8869            \\ \hline
\textbf{2stepQA}    & \textbf{0.9311}    & \textbf{0.9305}   \\ \hline
\end{tabular}
    \begin{tablenotes}
      \small
      \item The best performing algorithm is denoted in boldface.
    \end{tablenotes}
  \end{threeparttable}
\end{table}

\begin{table}[]
\centering
\begin{threeparttable}
\centering
\caption{Performance Comparison Between Different Combination of Two-Step R-NR IQA Models.}
\label{TS}
\begin{tabular}{|c|c|c|}
\hline
Reference + No-Reference          & SROCC           & PCC             \\ \hline
PSNR + NIQE                   & 0.6095          & 0.6181          \\ \hline
PSNR + BRISQUE               & 0.6868          & 0.6808          \\ \hline
\textbf{2stepQA}   & \textbf{0.9311} & \textbf{0.9305} \\ \hline
MS-SSIM + BRISQUE               & 0.9300          & 0.9277          \\ \hline
FSIM + NIQE                   & 0.9260          & 0.9290          \\ \hline
FSIM + BRISQUE                & 0.9280          & 0.9300          \\ \hline
VSI + NIQE                  & 0.8766          & 0.8833          \\ \hline
VSI + BRISQUE               & 0.8887          & 0.8919         \\ \hline
\end{tabular}
    \begin{tablenotes}
      \small
      \item The proposed 2stepQA Model (MS-SSIM + NIQE) is denoted by boldface.
    \end{tablenotes}
  \end{threeparttable}
\end{table}

We now describe the experimental analysis of 2stepQA and the other models that was carried out on the LIVE Wild Compressed Picture Database. The performances of the objective prediction models was measured using the Spearman Rank Order Correlation Coefficient (SROCC) and the Pearson Correlation Coefficient (PCC). The former measures the monotonicity between the subjective and objective scores, while the latter measures the degree of linear agreement between them. The subjective scores were processed by the subject rejection protocols in \cite{ITU} and \cite{li2017recover} which yield similar results. For both evaluation metrics, a larger value denotes better model performance. The reported correlation values were calculated by performing 1000 random 80\%-20\% splits without content overlap, then taking the median value of the results. Before computing LCC, a logistic non-linearity \cite{seshadrinathan2010study} was used to map the quality scores to the subjective scores.

\begin{figure}
\centering
\includegraphics[width = 0.65\columnwidth]{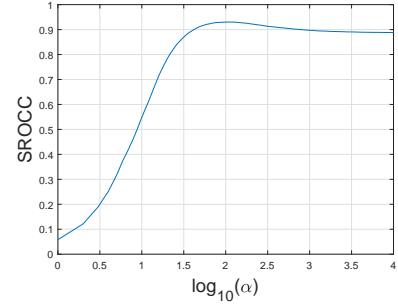}
\caption{The performance of 2stepQA index under varied $\alpha$ values.}
\label{alpha2}
\end{figure}

The rescaled indices (\ref{2step}) and (\ref{MN}) in each case were computed and these parameters were held constant in all the experiments. We compared 2stepQA against the state-of-the-art Reference models PSNR, MS-SSIM \cite{wang2003multiscale}, FSIM \cite{zhang2011fsim} and VSI \cite{zhang2014vsi}, with results given in Table \ref{alone}. Among the Reference models, PSNR was the worst-performing, but it does not utilize any perceptual properties. MS-SSIM and FSIM both delivered good performances. However, 2stepQA outperformed all of the other approaches showing that accounting for the quality of the reference source image is an effective strategy. We also report the performance of the NR models NIQE and BRISQUE (applied on the distorted images only), and found that both performed well, but not as well as 2stepQA. Fig. \ref{alpha2} plots the performance of (\ref{MN}) against $\alpha$. The 2stepQA index clearly attains its best performance at about $\alpha = 100$.

One of the greatest advantages of the 2stepQA framework is that it can incorporate Reference and/or No-Reference metrics depending on the application. To demonstrate this, we also report the results of other Reference-No-Reference combinations in Table \ref{TS} using (\ref{3ab}), (\ref{4R}) and (\ref{5NR}). The value of $\beta$ was determined to yield the best performance on each training set, then the predictor was evaluated on testing set. Note that when using BRISQUE as the NR model, the reference images in the training set were used to predict the quality of the reference images in the corresponding test set. Across all possible Reference-No-Reference combinations, we observed significant improvements when using the (\ref{2step}). The performance of the simple 2stepQA model (\ref{MN}) is remarkable given that it is applied without the need for any distortion-specific information or subjective data. 

\section{Future Work}
\label{5}
We proposed a novel two-step Reference-No-Reference IQA framework that integrates Reference and No-Reference information using a simple and efficient multiplication step. We evaluated the proposed approach on a recently developed subjective database and found that it produces standout performance compared to state-of-the-art Reference models. We envision developing more sophisticated bayesian approaches for integrating Reference and No-Reference information \cite{bovik201775} e.g., by conditioning Reference quality predictions on prior No-Reference predictions of the source picture. We believe that the ideas presented here relate fully to the video quality assessment (VQA) problems as well, in cases where the original source video is not of high quality and is subjected to authentic source inspection prior to compression. 

\bibliographystyle{IEEEtran}
\bibliography{bibfile}{}
\end{document}